# Characterization of the frequency response of channel-interleaved photonic ADCs based on the optical time-division demultiplexer

Na Qian, Linbo Zhang, Jianping Chen, and Weiwen Zou, * Senior Member, IEEE

State Key Laboratory of Advanced Optical Communication Systems and Networks, Intelligent Microwave Lightwave Integration Innovation Center (imLic), Department of Electronic Engineering, Shanghai Jiao Tong University, Shanghai 200240, China
*Email:  wzou@sjtu.edu.cn

**Abstract:** We characterize the frequency response of channel-interleaved photonic analog-to-digital converters (CI-PADCs) theoretically and experimentally. The CI-PADC is composed of a photonic frontend for photonic sampling and an electronic backend for quantization. The photonic frontend includes a photonic sampling pulse generator for directly high-speed sampling and an optical time-division demultiplexer (OTDM) for channel demultiplexing. It is found that the frequency response of the CI-PADC is influenced by both the photonic sampling pulses and the OTDM, of which the combined impact can be characterized through demultiplexed pulse trains. First, the frequency response can be divided into multiple frequency intervals and the range of the frequency interval equals the repetition rate of demultiplexed pulse trains. Second, the analog bandwidth of the CI-PADC is determined by the optical spectral bandwidth of demultiplexed pulse trains which is broadened in the OTDM. Further, the effect of the OTDM is essential for enlarging the analog bandwidth of the CI-PADC employing the photonic sampling pulses with a limited optical spectral bandwidth.

## 1. Introduction

Photonic analog-to-digital converters (PADC) are applied to receive and digitize high-speed microwave signal in next-generation radar and communication systems thanks to the superiorities of ultra-broad input bandwidth and ultra-high sampling rate [1-5]. Among various kinds of PADC schemes, the photonic sampled and electronic digitized PADC which takes advantage of photonics in wideband processing and that of electronics in high precision quantization has become one of the mainstream schemes [6-15]. In this kind of PADC scheme, the microwave signal is sampled by photonic sampling pulses and digitized by the electronic quantizers [6, 9]. Thanks to the high repetition rate of pulsed light sources, the high-speed photonic sampling can be directly realized without multiplexing [2, 12, 13]. Whereas the high-speed photonic sampling pulses need to be demultiplexed into multiple parallel channels to match the quantizing rate of the electronic quantizer in each channel. That is, the PADC with high-speed photonic sampling pulses and an optical time-division demultiplexer (OTDM) is classified as the channel-interleaved PADC (CI-PADC). The CI-PADC has been demonstrated in the photonics-based radar system and proven to be practical [2].

In order to meet the need of working at different frequency bands with different bandwidths in next-generation radar systems, the PADCs with flat frequency response and large analog bandwidth are necessary. Although the Nyquist bandwidth equals half of the total sampling rate, the frequency range of signal can be received is determined by the analog bandwidth. The analog bandwidth is the frequency range over which the signal power is attenuated by less than 3 dB of the maximum signal transmission [16]. It can be measured with the frequency response of the PADCs. Ref. [16] has verified the frequency response of single- channel PADCs only depends on the bandwidth of the electronic-optical modulator (EOM) and the temporal width of photonic sampling pulses. Benefitting from the development of optoelectronic devices, EOMs with ultra-large bandwidth have been reported [17]. However, the frequency response of CI-PADCs, which is complex with the influence of OTDM, has not been quantitatively characterized. Besides, high-speed photonic sampling pulses in CI-PADCs directly generated by cascaded modulators has a limited optical spectral bandwidth compared with mode-locked lasers [4, 10, 18]. The characterization of the frequency response is more essential for CI-PADCs employing photonic sampling pulses with narrower optical spectral bandwidth.

In this work, we theoretically and experimentally characterize the frequency response of CI-PADCs. By modeling the CI-PADC, the frequency response of CI-PADCs is revealed to be jointly affected by photonic sampling pulses and the OTDM, of which the combined impact can be characterized through demultiplexed pulse trains. The frequency responses of CI-PADCs with different photonic sampling pulses and different OTDMs are measured experimentally. The results indicate that the analog bandwidth of CI-PADCs is expanded through the OTDM when the optical spectral bandwidth of photonic sampling pulses is limited.

## 2. Principles

The schematic of the CI-PADC is illustrated in Fig. 1(a), including a photonic frontend and an electronic backend. In the photonic frontend, the high-speed photonic sampling pulses are directly generated by the photonic sampling pulse generator, which is based on mode-locked lasers or cascaded modulators [2, 10, 18]. Photonic sampling pulses sample the analog input signal at the sampling rate of $f_S$ when passing through the electronic-optical modulator (EOM). The photonic sampling pulses carrying the information of analog input signal are demultiplexed into $N$ parallel channels by the OTDM. The OTDM consists of cascaded dual-output Mach-Zehnder modulators (DO-MZMs). For each DO-MZM, the repetition rate of output pulse trains equals half of the input pulse trains as two adjacent input pulses pass through the DO-MZM at the maximal and the minimal transmittances [19]. With cascaded configuration, the repetition rate of pulse trains output from the OTDM is $f_S/N$ after passing through $log_2N$ DO-MZMs. Based on the DO-MZMs, the OTDM functions without any wavelength demultiplexer or dispersion element. The frequency of microwave signal input to each DO-MZM is equivalent to the repetition rate of pulse trains output from itself. As shown in Fig. 1(a), the microwave signal input to DO-MZMs works as the microwave driver which is directly divided from the clock of photonic sampling pulse generator. The frequency of microwave driver is $f_S/2$, $f_S/4$… $f_S/N$ and the amplitude of the microwave driver is adjusted to match the $V_\pi$ of DO-MZMs [19]. Then the photodetector (PD) in each channel converts demultiplexed pulse trains into electronic signals. In the electronic backend, the parallel electronic analog-to-digital converters (EADCs) quantize the electronic signals into digital series simultaneously. Finally, the digital signal processor (DSP) interleaves and reconstructs the digitized series from each channel to obtain the sampling rate of $f_S$.

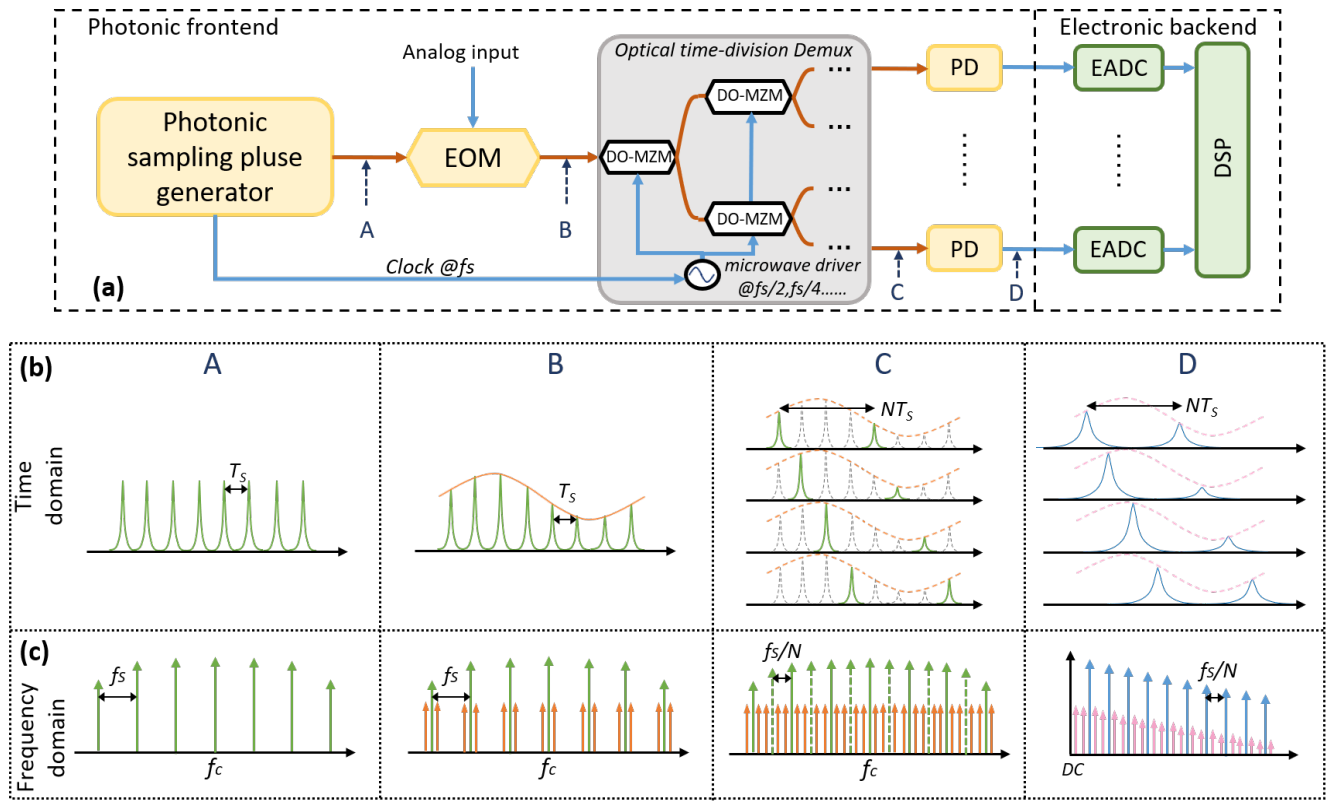

Fig. 1. (a) Architecture of the CI-PADC. EOM: electronic-optical modulator; DO-MZM: dual-output Mach-Zehnder modulator; PD: photodetector; EADC: electronic analog-to-digital converter; DSP: digital signal processor. (b) Schematic diagram of the CI-PADC in the time domain. (c) Schematic diagram of the CI-PADC in the frequency domain.

The high-speed photonic sampling pulses contain multiple comb lines and can be expressed as:

$$p(t)=\sum_{m=-L}^{L} p_m e^{j2\pi(f_C+mf_S)t}, \qquad (1)$$

where $p_m$ are coefficients that determine the power of different comb lines; $f_C$ is the spectrum central frequency; and $f_S$ is the comb lines spacing. The repetition rate of the photonic sampling pulses is also $f_S$. $L$ is an integer number and photonic sampling pulses has $2L+1$ comb lines. The waveforms both in the time domain and the frequency domain representing the output from the photonic sampling pulse generator (point A) are depicted in Fig. 1(b) and Fig. 1(c), respectively. Supposing the analog input signal is $v_{in}(t)=cos(2\pi f_0 t)$ at the frequency of $f_0$, $\alpha$ is the modulation index. After high-speed photonic sampling, the output of EOM (point B) can be derived as:

$$v(t)=(1+\alpha\cos(2\pi f_0 t))\sum_{m=-L}^{L} p_m e^{j2\pi(f_c+mf_s)t}$$
$$=\sum_{m=-L}^{L} p_m \left\{ e^{j2\pi(f_c+mf_s)t} + \frac{\alpha}{2}\left[ e^{j2\pi(f_c+mf_s+f_0)t} + e^{j2\pi(f_c+mf_s-f_0)t} \right] \right\}. \tag{2}$$

Equation (2) reveals that each comb line has an upper sideband and a lower sideband after photonic sampling, which are $+f_0$ and $-f_0$ away, respectively. Here, we assume that the EOM is in linear region and the amplitude of analog input signal is small.

As shown in Fig. 1(a), the period of photonic pulse trains output from the OTDM is $NT_S$, then the temporal response of OTDM can be regarded as periodic switching window. The repetition rate of this periodic switching window is $f_S/N$. After the Fourier series expansion, the temporal response of OTDM can be written as:

$$h(t)=\sum_{i=-D}^{D} b_i e^{j2\pi i \frac{f_s}{N} t}, \tag{3}$$

where $N$ is the number of demultiplexing channels; $b_i$ are Fourier series coefficients for different harmonics at $if_S/N$; $D$ is an integer number and $2D+1$ represents the number of harmonics. The value of $b_i$ and $D$ will be explained below. Combining Eq. (2) and Eq. (3), the output of each channel from the OTDM (point C) can be derived as:

$$v_n(t_n)=\sum_{m=-L}^{L}\sum_{i=-D}^{D} p_m b_i \left\{ e^{j2\pi(f_c+mf_s+\frac{if_s}{N})t_n} + \frac{\alpha}{2}\left[ e^{j2\pi(f_c+mf_s+\frac{if_s}{N}+f_0)t_n} + e^{j2\pi(f_c+mf_s+\frac{if_s}{N}-f_0)t_n} \right] \right\}$$
$$=\sum_{r=-NL-D}^{NL+D} a_r \left\{ e^{j2\pi(f_c+\frac{rf_s}{N})t_n} + \frac{\alpha}{2}\left[ e^{j2\pi(f_c+\frac{rf_s}{N}+f_0)t_n} + e^{j2\pi(f_c+\frac{rf_s}{N}-f_0)t_n} \right] \right\}, \tag{4}$$

where $t_n=t-(n-1)T_S$. It represents that the time interval between adjacent channels is $T_S$. Equation (4) reveals that each original comb line at $f_C+mf_S$ has $2D$ generated comb lines around itself, as shown in Fig. 1(c). The spacing of all comb lines including original comb lines and generated comb lines in $v_n(t_n)$ is $f_S/N$. These comb lines are weighted by $a_r$, which is determined by the optical spectrum of photonic sampling pulses and the response of OTDM. The optical spectral bandwidth of demultiplexed pulse trains is broadened from $2Lf_S$ to $2f_S(L+D/N)$. $v_n(t_n)$ contains $2\times[2(NL+D)+1]$ replicas of $v_{in}(t)$ at frequencies $f_C+rf_S/N\pm f_0$. These replicas are also weighted by $a_r$. After the photodetection, the output from the PD in each channel (point D) is expressed as:

$$r_n(t_n)=R_0\eta^2 |v_n(t_n)|^2$$
$$=R_0\eta^2 \sum_{k=0}^{2(NL+D)} \sum_{r=-NL-D}^{NL+D-k} a_r a_{r+k} \left\{ e^{j2\pi \frac{kf_s}{N} t_n} + \alpha\left[ e^{j2\pi(\frac{kf_s}{N}+f_0)t_n} + e^{j2\pi(\frac{kf_s}{N}-f_0)t_n} \right] \right\}, \tag{5}$$

where $R_0$ and $\eta$ are the load resistance and the responsivity of the PD, respectively. Here, we consider all terms given by line-lines and line-sidebands beatings of $v_n(t_n)$ in PDs. We ignore other beat products which are usually much weaker [5]. From Eq. (5) and Fig. 1(c), it can be concluded that the harmonics at $kf_S/N$ and the replicas of $v_{in}(t)$ at frequencies $kf_S/N\pm f_0$ are weighted by the convolution result of Fourier series coefficients $a_r$. Then, we define a new coefficient $g_k$ as:

$$g_k=\sum_{r=-NL-D}^{NL+D-k} a_r a_{r+k}, \tag{6}$$

where $k$ is the integer number from 0 to $2(NL+D)$. Equation (5) can be rewritten as:

$$r_n(t_n)=R_0\eta^2 \sum_{k=0}^{2(NL+D)} g_k \left\{ e^{j2\pi \frac{kf_s}{N} t_n} + \alpha\left[ e^{j2\pi(\frac{kf_s}{N}+f_0)t_n} + e^{j2\pi(\frac{kf_s}{N}-f_0)t_n} \right] \right\}. \tag{7}$$

If the bandwidth of the PD is ultrabroad, the PD output is a series of harmonics and replicas of $v_{in}(t)$ weighted by $g_k$. According to Eq. (4) ~ Eq. (6), the value of $g_k$ is determined by $p_m$ and $b_i$. In other words, photonic sampling pulses and the OTDM affect the power of harmonics at $kf_S/N$ and replicas of $v_{in}(t)$ jointly. In fact, the bandwidth of PDs and the EADCs is limited, which is set as $f_S/2N$, namely the Nyquist bandwidth of the demultiplexing channels. Therefore, only the replica of $v_{in}(t)$ within the frequency range $[0, f_S/2N]$ can be selected and digitized. For simplicity, the frequency response of the PD, $H_{PD}(f)$, can be approximately described as a low-pass filter, which is a rectangular model of $H_{PD}(f) = 1$ when $|f| \leq f_S/2N$ and $H_{PD}(f) = 0$ when $|f| > f_S/2N$ [20, 21]. The digitized output from each EADC can be expressed in the frequency domain as:

$$V_n[f]=A_n(f)g_{k_0}\delta\left[ f-\left|\frac{k_0 f_s}{N}-f_0\right| \right], \tag{8}$$

where the value of $A_n(f)$ is related to the amplitude of the analog input signal, frequency response of the EOM, the load resistance $R_0$, and the responsivity $\eta$ of the PD. Here, to concentrate on the mechanism of photonic sampling pulses and the OTDM, the value of $A_n(f)$ is taken as a constant. $k_0$ is the integer satisfying $-f_S/2N < k_0 f_S/N - f_0 < f_S/2N$. It can be found that the analog input signal $v_{in}(t)$ in different frequency intervals is weighted by different $g_k$. After channel interleaving through DSP, the digitized output from each channel is reconstructed to achieve the sampling rate of $f_S$. Then the output of the CI-PADC is calculated as:

$$V_{out}[f]=\begin{cases} A_n(f)g_{k_0}\delta\left[f-\left|k_0'f_S-f_0\right|\right], & |f|\leq f_S/2 \\ 0, & |f|>f_S/2 \end{cases} \quad (9)$$

where $k_0'$ is the integer satisfying $-f_S/2 < k_0'f_S - f_0 < f_S/2$. Comparing Eq. (9) with Eq. (8), channel interleaving in DSP only changes the position of output signal in the electrical spectrum. This is because that the power of the signal is fixed and will not be amplified or attenuated by the channel interleaving once it is output from EADCs. The power of the output signal in different frequency intervals is weighted by different $g_k$, which is determined by photonic sampling pulses and the OTDM. The range of each frequency interval is $f_S/N$ according to the criteria for selecting $g_k$ ($-f_S/2N < k_0f_S/N-f_0 < f_S/2N$). Based on the above analysis, it is essential to quantize the value of $g_k$ for the characterization of frequency response of CI-PADCs.

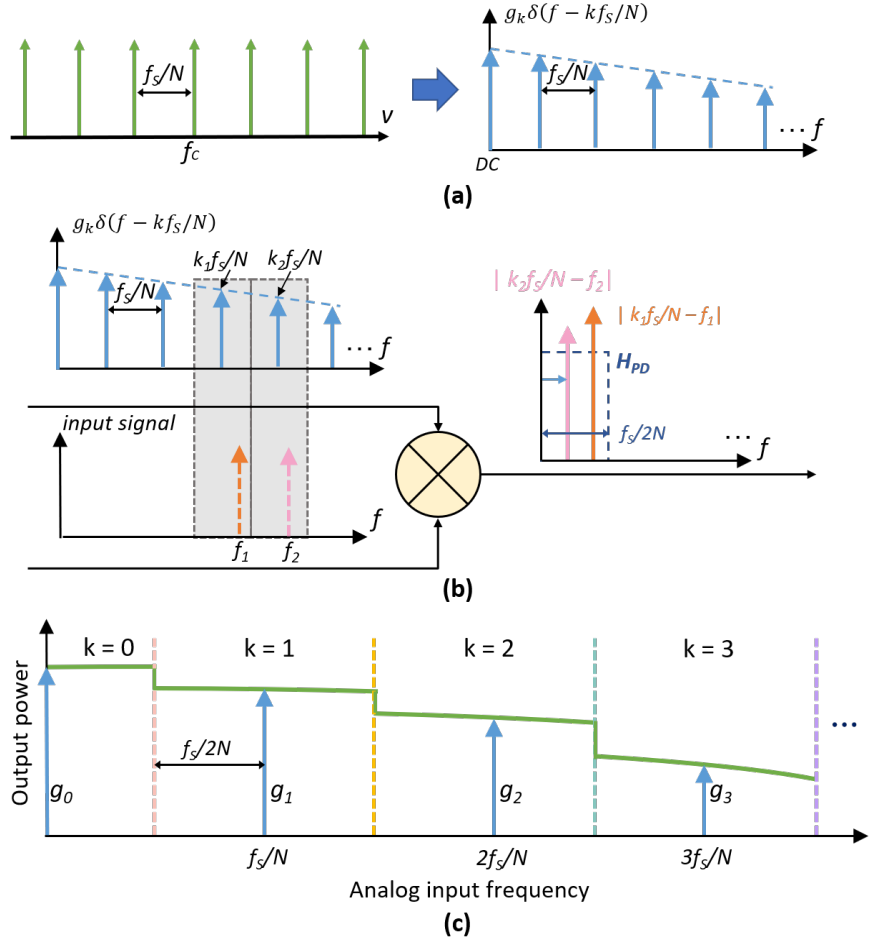

Fig. 2. (a) Schematic of line-lines beatings of original and generated comb lines in demultiplexed pulse trains. (b) Spectral schematic of the CI-PADC. (c) Frequency response of the CI-PADC under the influence of photonic sampling pulses and the OTDM.

When the analog signal into CI-PADCs is absent, the output from each channel in the frontend is line-lines beatings of original and generated comb lines in demultiplexed pulse trains, spaced at $f_S/N$, as shown in Fig. 2(a). Its result can be expressed as:

$$r_{demux}(t_n)=R_0\eta^2\sum_{k=0}^{2(NL+D)}\sum_{r=-NL-D}^{NL+D-k}a_ra_{r+k}e^{j2\pi\frac{kf_S}{N}t_n}=R_0\eta^2g_ke^{j2\pi\frac{kf_S}{N}t_n}. \quad (10)$$

After photodetection, the power of demultiplexed pulse trains on the electrical spectrum is also weighted by $g_k$. Fourier transform of Eq. (10) is:

$$V_{demux}(f)=g_k\delta(f-kf_S/N). \quad (11)$$

Here, we ignore the effects of $R_0$ and $\eta$ of the PD since they are constants. Combining Eq. (7) − Eq. (11), the output of CI-PADCs is equivalent to the result of mixing between analog input signal and demultiplexed pulse trains in the frequency domain, as depicted in Fig. 2(b). The output power of the CI-PADC is weighted by the $g_{k0}$, which is the power of the harmonic $k_0f_S/N$ closest to the frequency $f_0$ of analog input signal. As illustrated in Fig. 2(b), assuming there is a dual-tone analog input signal $v_{in2}(t)=cos(2\pi f_1t) + cos(2\pi f_2t)$, $f_1$ and $f_2$ satisfy $-f_S/2N < k_1f_S/N-f_1 < f_S/2N$ and $-f_S/2N < k_2f_S/N-f_2 < f_S/2N$, respectively. The output power of the signal at $f_1$ and $f_2$ are weighted by $g_{k1}$ and $g_{k2}$, respectively. Derived from Eq. (9) and

Fig. 2(c), the frequency response of the CI-PADC determined by photonic sampling pulses and the OTDM is stepped distribution. The frequency range of each step is $f_s/N$. The flatness of the frequency response depends on the value of $g_k$. This effect is different from the frequency response affected by the EOM.

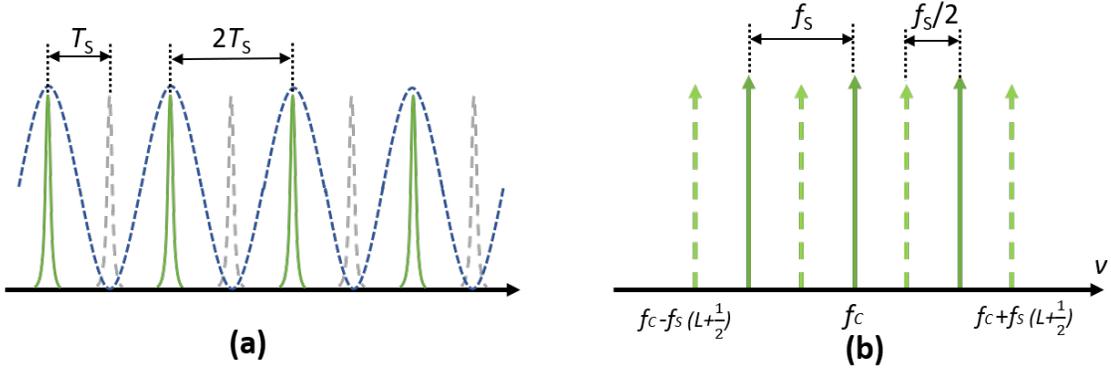

Fig. 3. Demultiplexed pulse trains output from the OTDM with single DO-MZM in (a) time domain and (b) frequency domain.

According to Eq. (6), $g_k$ is the convolution result of Fourier series coefficients $a_r$ and $a_r$ represents the power of comb lines in demultiplexed pulse trains. The value of $a_r$ can be calculated through Eq. (4) and the temporal response of OTDM. OTDM consists of cascaded DO-MZMs. Therefore, we further investigate the temporal response of OTDM via the temporal response of single DO-MZM [22], which is expressed as:

$$h_a(t) = \frac{\alpha_{max}}{2}\left[\left(1+\frac{1}{\varepsilon}\right)+\left(1-\frac{1}{\varepsilon}\right)\mu\cos(2\pi f_{driver}t)\right], \quad (12)$$

where $h_a(t)$ is the temporal response of single DO-MZM. $\varepsilon$ is the optical extinction ratio of the DO-MZM and can be calculated by $\varepsilon=\alpha_{max}/\alpha_{min}$. $\alpha_{max}$ and $\alpha_{min}$ are the maximum and minimum transmittances respectively. $\mu$ is the index used to characterize the performance of DO-MZM which can be adjusted to its maximum value according to Ref. [22, 23]. The maximum value of $\mu$ is 1. $f_{driver}$ is the frequency of microwave driver into the DO-MZM. For the OTDM with single DO-MZM, the number of demultiplexing channels $N=2$, $f_{driver}=f_s/2$, $\mu=1$. The optical extinction ratio $\varepsilon$ is extremely high and the influence of $1/\varepsilon$ can be neglected. Equation (12) can be rewritten as:

$$h_a(t) = \frac{\alpha_{max}}{2}\left[1+\frac{1}{2}\left(e^{j2\pi\frac{f_s}{2}t}+e^{-j2\pi\frac{f_s}{2}t}\right)\right], \quad (13)$$

Here, the values of $b_i$ and $D$ in Eq. (3) are made clear. The demultiplexed pulse trains output from the OTDM with single DO-MZM are illustrated in Fig. 3(a) and 3(b) in the time domain and the frequency domain, respectively. Combining Eq. (1), Eq. (4), and Eq. (13), the demultiplexed pulse trains can be derived as:

$$p_{demux}(t) = \sum_{r=-2L-1}^{2L+1} a'_r e^{j2\pi(f_c+r\frac{f_s}{2})t}. \quad (14)$$

There are 2(2L+1) +1 comb lines in demultiplexed pulse trains. The optical spectrum bandwidth of demultiplexed pulse trains is $(2L+1)f_s$ and broader than the original optical spectrum bandwidth of photonics sampling pulses $(2Lf_s)$. $a'_r$ is the power of comb lines in demultiplexed pulse trains. When the power of original comb lines $p_m$ are the same, based on Eq. (4) and Eq. (13), the values of $a'_r$ are equal when $r \neq \pm(2L+1)$. The value of $a'_r$ for $r = \pm(2L+1)$ is half of the value of $a'_r$ for $r \neq \pm(2L+1)$. It means the values of most $a'_r$ are equal. Then, we approximate all $a'_r$ to be $a'$ and the values of $g_k$ for $N=2$ can be calculated as: $g_k = a'^2[2(2L+1)+1-k]$. We define $M = 2(2L+1)+1$, where $M$ is the number of all comb lines in demultiplexed pulse trains. $g_k$ is rewritten as: $g_k = a'^2(M-k)$. $g_0 = a'^2M$, which is maximum in all values of $g_k$. Applying the value of $g_k$ into Eq. (9), the $k$ which is the minimum integer satisfying $g_k/g_0 \geq 1/2$, determines the analog bandwidth of this two-channel CI-PADC within 3 dB frequency response. It can be concluded that the analog bandwidth is larger when there are more comb lines in the demultiplexed pulse trains. In other words, when the comb line spacing is fixed, the more comb lines the broader optical spectrum bandwidth. In fact, the comb line spacing in the demultiplexed pulse trains equals the quantizing rate of EADC in each channel, which is set to be fixed in CI-PADCs.

With cascaded configuration of DO-MZMs, more comb lines are generated in demultiplexed pulse trains. The values of $g_k$ for CI-PADCs with more demultiplexing channels can be derived through multiplying Eq. (13). The optical spectra of demultiplexed pulse trains in CI-PADCs with single channel ($N=1$), two channels ($N=2$), and four channels ($N=4$) are depicted in Fig. 4(a). $g_{1k}$, $g_{2k}$, and $g_{4k}$ are the values of $g_k$ when $N=1$, 2, and 4, respectively. The number of all comb lines in these three CI-PADCs are equal so as to maintain the same analog bandwidth. The solid comb lines in different colors represent the original comb lines in photonic sampling pulses and the dashed comb lines in different colors represent the comb lines generated by the OTDM. It can be seen from Fig. 4(a) that the requirement for the number of original comb lines in the photonic sampling pulses is different for these three CI-PADCs, which is reduced in the CI-PADC with more demultiplexing channels. In Fig. 4(a), under different channel numbers, the value of $f_S/N$ determined by the quantizing rate of EADC is equal and the sampling rate of the CI-PADC is increased through configuring more demultiplexing channels. Demultiplexing channels formed by the cascaded DO-MZMs in the OTDM not only increase the sampling rate of the CI-PADC, but also reduce the demand for the number of original comb lines in the photonic sampling pulses while maintaining a large analog bandwidth.

Figure 4(b) compares the values of $g_k$ in the single-channel PADC and the two-channel CI-PADC with the same sampling rate $f_S$ and the same optical spectral bandwidth of photonic sampling pulses. In the two-channel CI-PADC (N =2), the optical spectral bandwidth of demultiplexed pulse trains is broadened from $2Lf_S$ to $(2L+1)f_S$ after the OTDM with a DO-MZM. The generated comb lines make the value of $g'_{2k}$ in the two-channel CI-PADC closer in comparison to $g'_{1k}$ in the single-channel PADC. Then, the analog bandwidth of the two-channel CI-PADC is widened. According to Fig. 2 and Fig. 4, the frequency response of the CI-PADC is under the common influence of the photonic sampling pulses and the OTDM, which can be characterized by the electrical spectrum characteristics of the demultiplexed pulse trains.

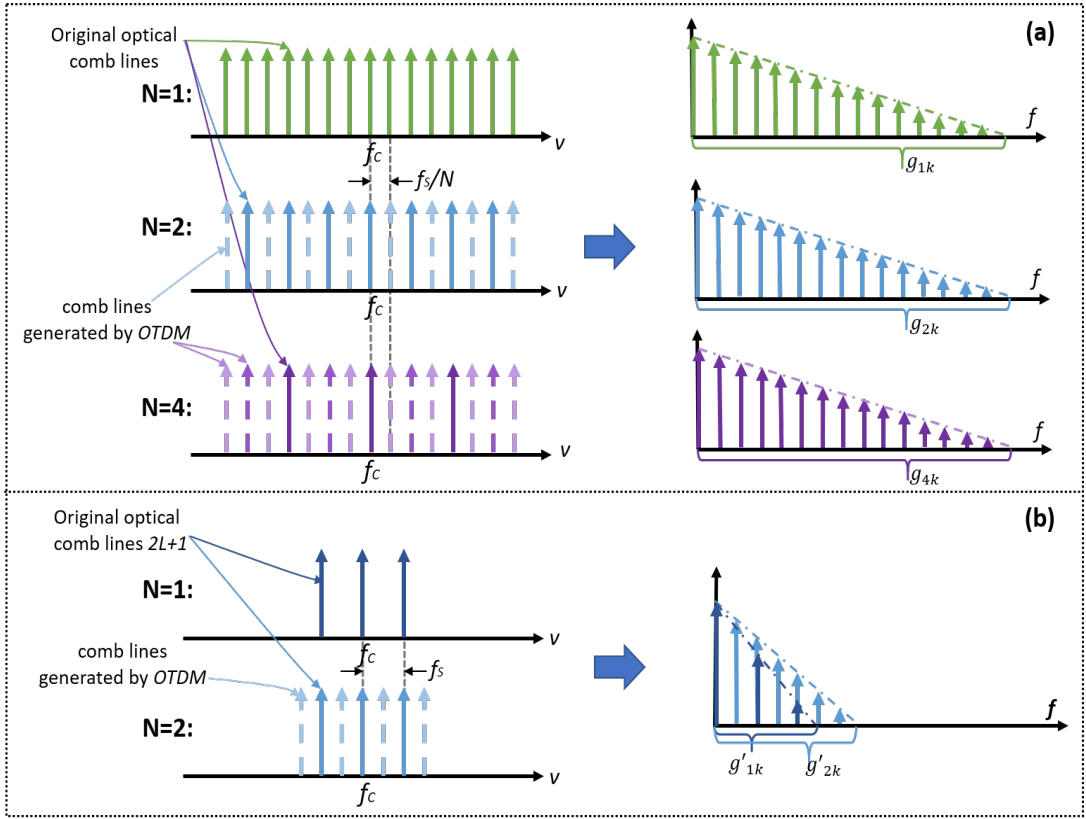

Fig. 4. (a) Optical spectra of demultiplexed pulse trains in the single-channel PADC, the two-channel CI-PADC, and the four-channel CI-PADC and their corresponding electrical spectra after photodetection. (b) Optical spectra of demultiplexed pulse trains in the single-channel PADC and the two-channel CI-PADC as well as their corresponding electrical spectra after photodetection when the sampling rate and the optical spectral bandwidth of photonic sampling pulses are same.

## 3. Experimental results

The experimental setups of the single-channel PADC, the two-channel CI-PADC and the four-channel CI-PADC are illustrated in Fig. 5. The main difference among these three PADCs is the OTDM. There is no OTDM in the single-channel PADC and the output from the EOM (PHOTLINE MXAN-LN-40, bandwidth= 40 GHz) is directly input to the PD array. The OTDM in the two-channel CI-PADC is composed of a DO-MZM (EOSpace AX-1x2-0MSS-20, bandwidth= 20 GHz) and the microwave driver at 10 GHz. The OTDM in the four-channel CI-PADC is composed of three DO-MZMs (EOSpace AX-1x2-0MSS-20, bandwidth= 20 GHz) and the microwave driver is at 20 GHz and 10 GHz. Another difference among these three PADCs is the photonic sampling pulse generator, which includes an actively mode-locked laser (AMLL, Calmar PSL-10-TT in single-channel PADC and the two-channel CI-PADC, Calmar PSL-40-TT in the four-channel CI-PADC) and the optical bandpass filter (OBPF, Alnair Labs, CVF-220CL). The repetition rate of the AMLLs is equivalent to the sampling rate of these three PADCs which are listed in Table 1. The timing jitter of the AMLLs is below 20 fs [21]. The OBPF is added to control the original comb lines in photonics sampling pulses. The EDFA in CI-PADCs is used to compensate for the loss caused by the OTDM. The number of PDs (ConquerLtd. KG-PT-10-SM-FA) and EADCs (Oscilloscope, Keysight MSOS804A) in the PD array and EADC array equals the number of demultiplexing channels in these three PADCs. The bandwidth and responsivity of the PDs in our experiment is 10 GHz and 0.9 A/W @1550 nm, respectively. The quantizing rate of EADCs is 10 GSa/s and quantization bits is 10 bits. The bandwidth of EADCs working at 10 GSa/s is ~4.2 GHz and the ENOB of EADCs can be up to 8.7 bits [24]. The DSP is realized by a computer to interleave the digital series from EADCs in each channel. AMLLs, microwave drivers, and EADCs are all kept synchronized.

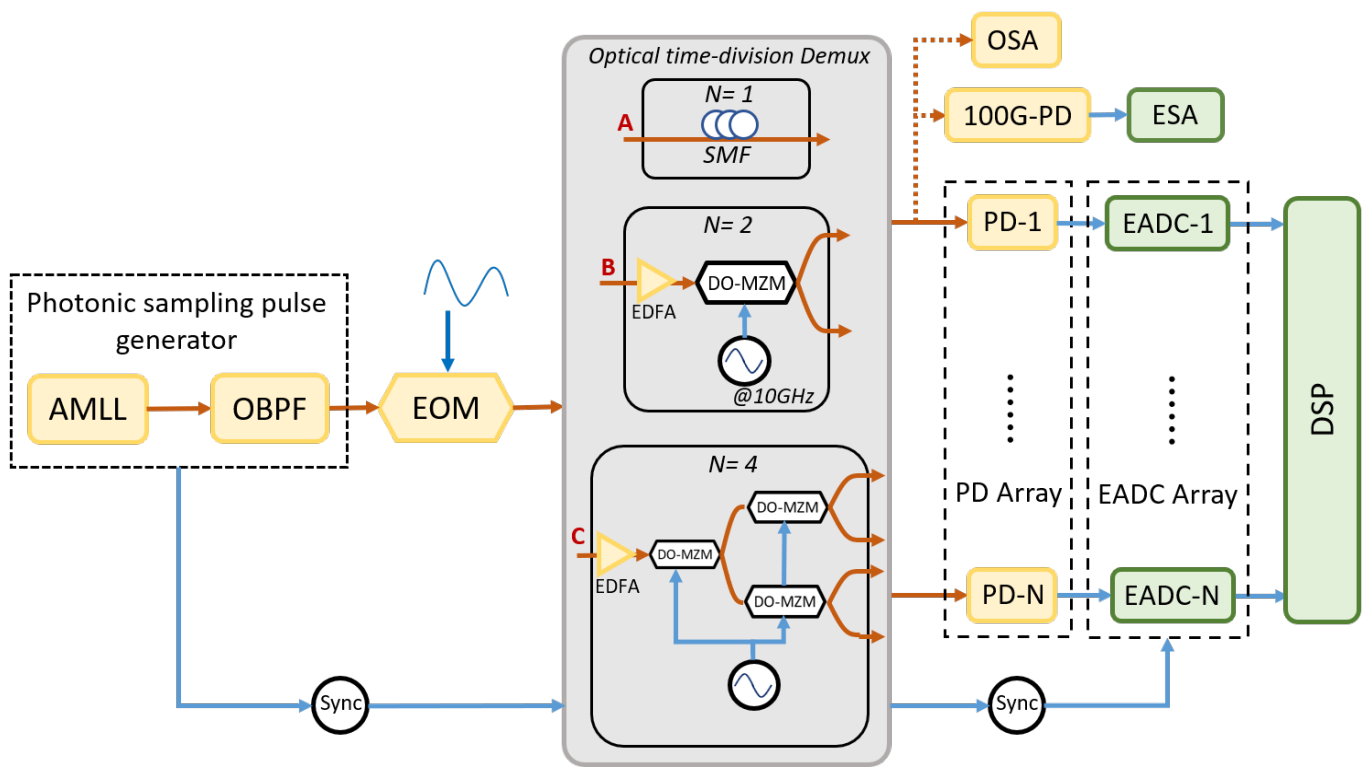

Fig. 5. Experimental setups of the single-channel PADC, the two-channel CI-PADC and the four-channel CI-PADC. AMLL: actively mode-locked laser; OBPF: optical bandpass filter; EOM: electronic-optical modulator; DO-MZM: dual-output Mach-Zehnder modulator; SMF: single mode fiber; EDFA: erbium-doped fiber amplifier; PD: photodetector; ESA: electrical spectrum analyzer; OSA: Optical spectrum analyzer; EADC: electronic analog-to-digital converter; DSP: digital signal processor; Sync: synchronization.

The optical spectra of photonic sampling pulses output from the OBPF in the three PADCs are measured by the optical spectrum analyzer (OSA, Yokogawa AQ6370C). As shown in Fig. 6(a), we filter out photonic sampling pulses with 3 lines, 7 lines and 15 lines in the single-channel PADC. The interval between adjacent original comb lines is 0.08nm. In the two-channel CI-PADC (Fig. 6(c)), we filter out photonic sampling pulses with 3 lines, 7 lines and the interval between adjacent original comb lines is 0.16nm. In the four-channel CI-PADC (Fig. 6(f)), we only filter out photonic sampling pulses with 3 lines and the interval between adjacent original comb lines is 0.32nm. The optical spectra of photonic sampling pulses in the three PADCs are obviously different. As depicted in Fig. 5, the optical spectra of demultiplexed pulse trains output from the OTDM in the three PADCs are also measured by the OSA. In the single-channel PADC, photonic sampling pulses are not demultiplexed. In the two-channel CI-PADC (Fig. 6(d)), there are 7 lines and 15 lines in the demultiplexed pulse trains after the OTDM when the photonic sampling pulses with 3 lines and 7 lines, respectively. As illustrated in Fig. 6(g), there are 15 lines in the demultiplexed pulse trains after the OTDM when the photonic sampling pulses with 3 lines in the four-channel CI-PADC. Comparing Fig. 6(a) and Fig. 6(d), the optical spectra of demultiplexed pulse trains in the two-channel CI-PADC are similar to the optical spectra of photonic sampling pulses with 7 lines and 15 lines in the single-channel PADC. Comparing Fig. 6(a) and Fig. 6(g), the optical spectra of demultiplexed pulse trains in the four-channel CI-PADC are similar to the photonic sampling pulses with 15 lines in the single -channel PADC. The interval between adjacent comb lines in demultiplexed pulse trains is 0.08nm, as shown in Fig. 6(d) and Fig. 6(g).

The electrical spectra of demultiplexed photonic pulses output from the OTDM in the three PADCs with different photonic sampling pulses are measured by the electrical spectrum analyzer (ESA, Rohde & Schwarz, FSUP 50) via the PD (Finisar, XPDV4121R, bandwidth= 100 GHz). Figures 6(b), 6(e), and 6(h) correspond to the electrical spectra of single-channel PADC, two-channel CI-PADC, and four-channel CI-PADC, respectively. The interval between adjacent harmonics is 10 GHz. In the single -channel PADC, the power of harmonics at 20 GHz, 30 GHz, and 40 GHz increases with the number of comb lines in photonic sampling pulses. The power of harmonics at 10 GHz, 20 GHz, 30 GHz, and 40 GHz corresponds to the value of $g_k$ in Eq. (9) when $k$=1, 2, 3 and 4. In the single-channel PADC, the power of these four harmonics is similar when there are 15 lines in photonic sampling pulses. The power of these four harmonics in the two-channel CI-PADC with 7 lines and the four-channel CI-PADC with 3 lines in the original photonic sampling pulses is also similar.

TABLE 1

The number of demultiplexing channels, quantizing rate and sampling rate of PADCs.

|  | Number of channels | Quantizing rate GSa/s | Sampling rate GSa/s |
|---|---|---|---|
| single-channel PADC | 1 | 10 | 10 |
| two-channel CI-PADC | 2 | 10 | 20 |
| four-channel CI-PADC | 4 | 10 | 40 |

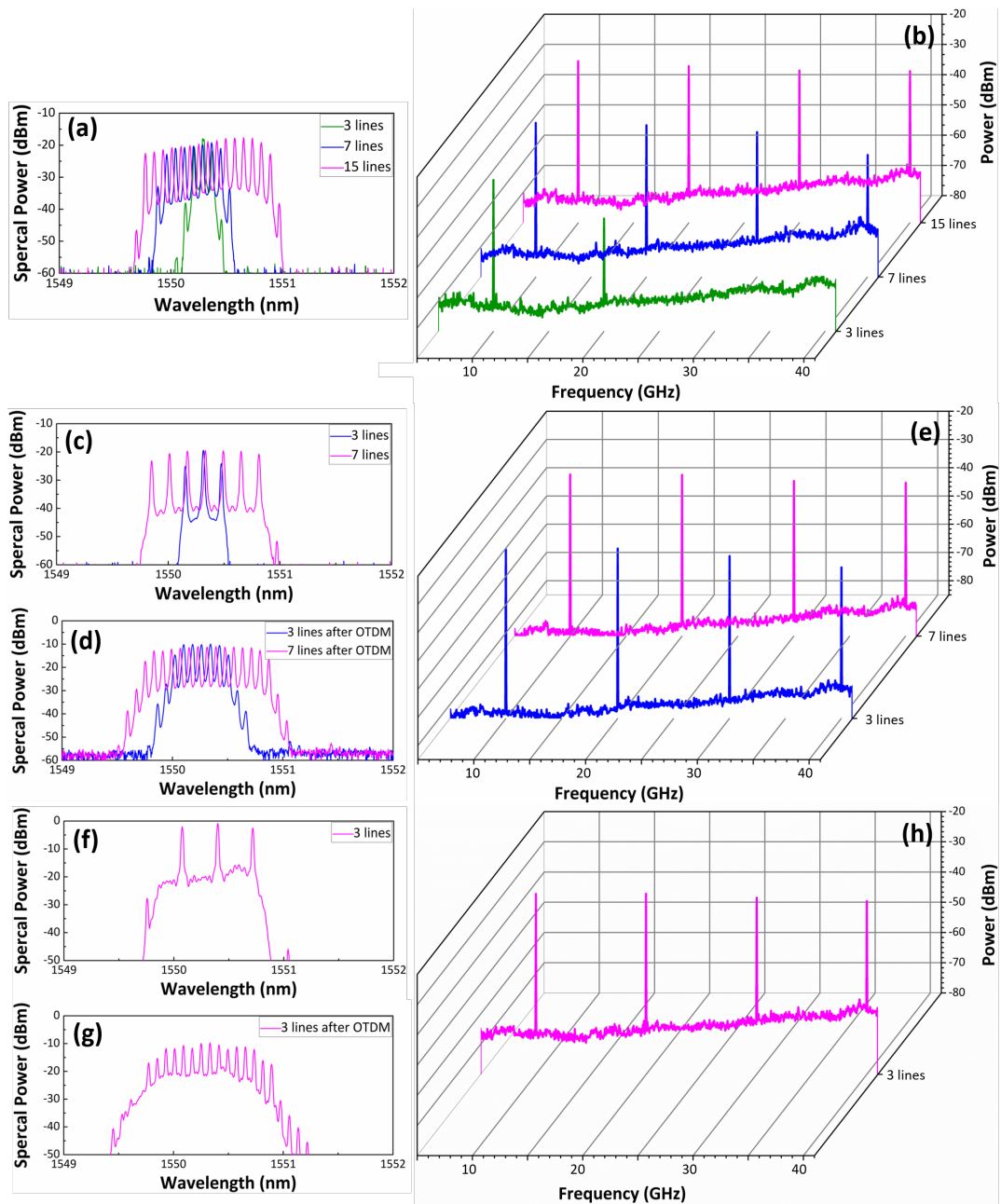

Fig. 6. The optical spectra of photonic sampling pulses output from the OBPF in (a): the single-channel PADC; (c): the two-channel CI-PADC and (f): the four-channel CI-PADC. The optical spectra of demultiplexed pulse trains output from the OTDM in (d): the two-channel CI-PADC and (g): the four-channel CI-PADC. The measured electrical spectra of demultiplexed photonic pulses in (b): the single-channel PADC; (e): the two-channel CI-PADC and (h): the four-channel CI-PADC.

    The frequency response of three PADCs with different photonic sampling pulses are experimentally acquired through inputting the analog signal from 0.5 GHz to 43.5 GHz. The output power of the signal at different input frequency is measured and calculated. Note that the influence of the EOM on the frequency response has been calibrated. The experimental results (symbols) of the three PADCs are displayed in Fig. 7 and compared with the theoretical estimations (lines). Theoretical estimations are calculated according to Eq. (9) and the value of $g_k$ from Fig. 6. Figure 7 reveals that the input signal in the range of $-f_S/2N < f_0 - k_0 f_S/N < f_S/2N$ ($f_S/N$ = 10 GHz) depends on the harmonics at frequency of $k_0 f_S/N$,

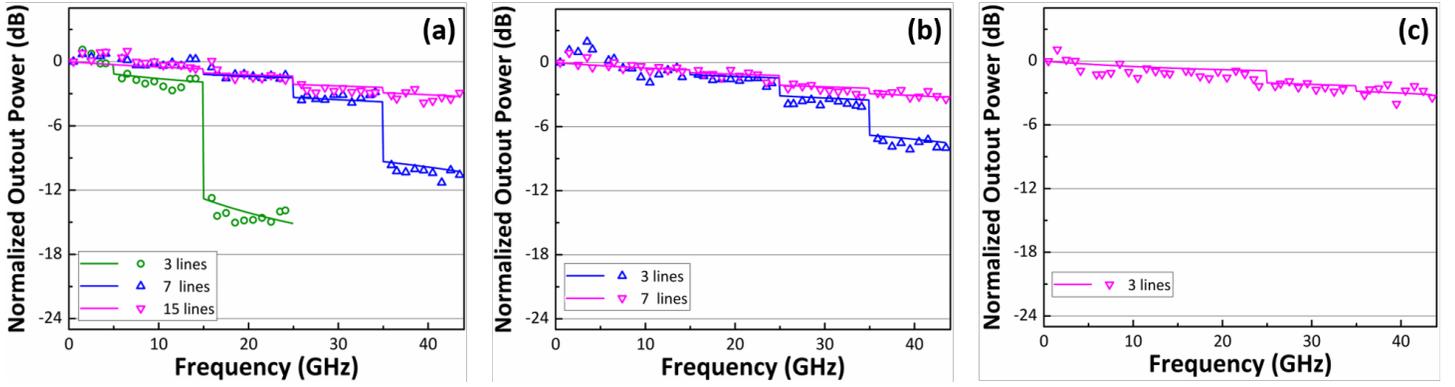

Fig. 7. The measured frequency response of PADCs with different photonic sampling pulses. (a): the single-channel PADC; (b): the two-channel CI-PADC; (c): the four-channel CI-PADC.

matching well with the theoretical analysis. As shown in Fig. 6, the power of the harmonics at $k_0 f_S/N$ ($k_0$= 1, 2, 3, 4 in the experiment) are different, then the measured frequency responses in the three PADCs is stepped. In the single-channel PADC, the power of different harmonics gets closer when the number of comb lines increases. In the case of 3 lines as depicted in Fig. 7(a), the analog bandwidth of the single-channel PADC is only ~15 GHz and the output power in the frequency range of 25~43.5 GHz is too low to be detected. As the number of optical comb lines increases, the analog bandwidth is observably enlarged. In CI-PADCs, there are comb lines generated by the OTDM in the demultiplexed pulse trains, the frequency response is determined by the photonic sampling pulses and the OTDM jointly. In Fig. 7(b), the analog bandwidth of the two-channel CI-PADC is ~35 GHz and over 40 GHz when the photonic sampling pulses with 3 lines and 7 lines, respectively. In Fig. 7(c), the analog bandwidth of the four-channel CI-PADC is over 40 GHz when the photonic sampling pulses with 3 lines. It is obvious that the frequency responses of the single-channel PADC with 15 lines, the two-channel CI-PADC with 7 lines, and the four-channel CI-PADC with 3 lines in photonics sampling pulses are almost the same. It is concluded that the OTDM can reduce the demand for the number of original comb lines in the photonic sampling pulses when maintains a large analog bandwidth. The experimental results are consistent with the theoretical analysis in Fig. 4.

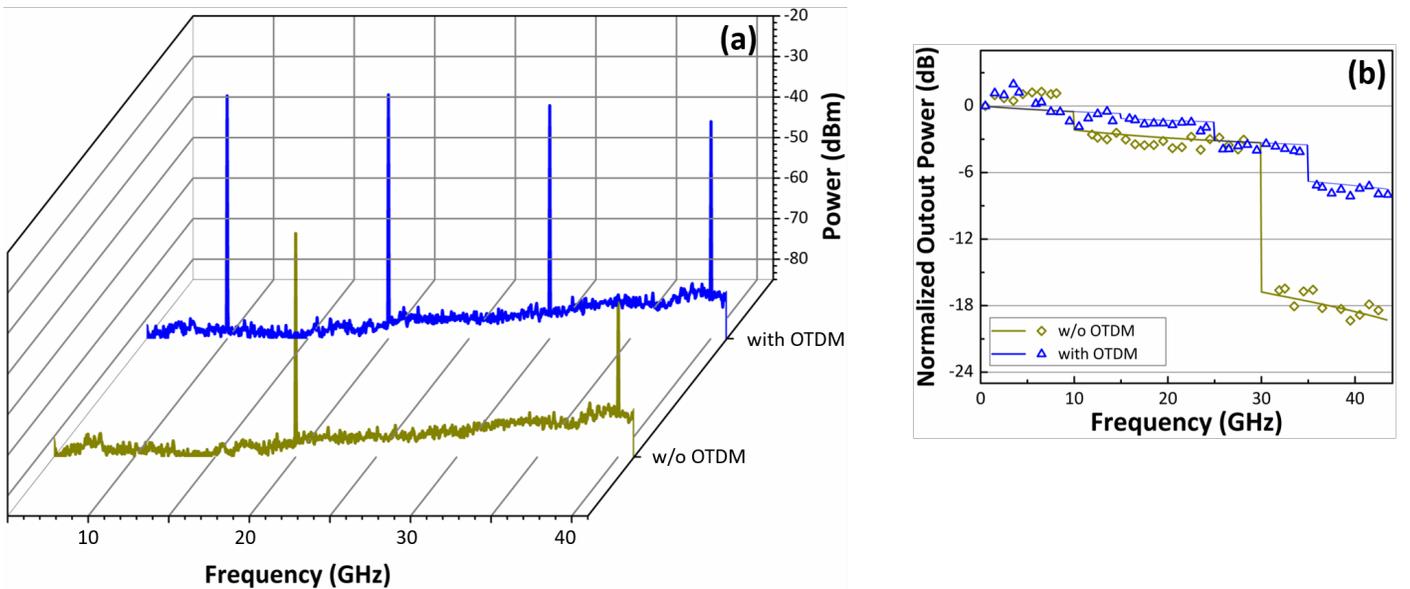

Fig. 8. Comparison of the 20GSa/s-PADC with and without an OTDM. (a): Electrical spectra of demultiplexed photonic pulses in the 20GSa/s-PADC with an OTDM and that of photonic sampling pulses in the 20GSa/s-PADC without an OTDM. (b): The measured frequency response of 20GSa/s-PADC with and without an OTDM.

We also experimentally compare frequency responses of the single-channel PADC and the two-channel CI-PADC with the same sampling rate $f_S$ ($f_S$ = 20GSa/s). The setups of the single-channel PADC and the two-channel CI-PADC are the same to Fig. 5. Here, the repetition rate of the AMLL (Calmar PSL-10-TT) employed in these two PADCs is 20 GHz which equals the sampling rate $f_S$. Controlled by the OBPF, the optical spectra of photonics sampling pulses in these two PADCs are the same. There are 3 comb lines in the photonics sampling pulses and the interval between adjacent comb lines is 0.16nm. The quantizing rate and bandwidth of EADC in the single-channel PADC is 20 GSa/s and 8.4GHz [24]. In the two-channel CI-PADC, the repetition rate of demultiplexed photonic pulses is 10 GHz. The quantizing rate and bandwidth of

EADC is 10 GSa/s and 4.2 GHz. The electrical spectrum of the photonic sampling pulses in the single-channel PADC is shown in Fig. 8(a) (brown). The interval between adjacent harmonics is 20 GHz. The electrical spectrum of demultiplexed photonic pulses in the two-channel CI-PADC is also shown in Fig. 8(a) (blue). The interval between adjacent harmonics is 10 GHz. It reveals that the comb lines generated by the OTDM change the interval and power difference between each harmonic. Accordingly, the frequency responses of these two PADCs are also different. As depicted in Fig. 8(b), with the effect of the OTDM, the frequency responses of the two-channel CI-PADC is flatter than the single-channel PADC. The analog bandwidth of the single-channel PADC is ~30 GHz and that of the two-channel CI-PADC is ~35 GHz. The results confirm the analog bandwidth of CI-PADCs employing photonic sampling pulses with a limited optical spectral bandwidth is enlarged by the OTDM.

## 4. Conclusions

The frequency response of the CI-PADC affected by the photonic sampling pulses and the OTDM has been investigated theoretically and experimentally. After theoretical modelling, it is found that the frequency response of the CI-PADC can be characterized through demultiplexed pulse trains and the optical spectrum bandwidth of demultiplexed pulse trains is expanded in the OTDM. The frequency responses of CI-PADCs with different photonic sampling pulses and different OTDMs have been measured. Experimental results indicate that the OTDM can reduce the demand for the number of original comb lines in the photonic sampling pulses while maintaining a large analog bandwidth. Our research can provide guidance for the design of on-chip photonic sampling pulses in future integrated CI-PADCs. It is expectable that the complexity and power consumption of the photonic sampling pulses generator in CI-PADCs can be further reduced.

## Acknowledgements

This work was supported in part by the National Key Research and Development Program of China (Program No. 2019YFB2203700) and the National Natural Science Foundation of China (Grant No. 61822508). The authors wish to thank the anonymous reviewers for their valuable suggestions.